\documentclass[prd,nofootinbib,english,superscriptaddress]{revtex4}

\usepackage[letterpaper,top=2cm,bottom=2cm,left=2.35cm,right=2.35cm,marginparwidth=1.75cm]{geometry}

\usepackage{graphicx,float}
\usepackage{amsmath,amssymb,amsfonts}
\usepackage{upgreek}
\usepackage{mathrsfs}
\usepackage{epsfig,color}
\usepackage[thinlines]{easytable}
\usepackage{pdfpages}
\usepackage{array}
\usepackage{cancel}
\usepackage{mathtools}
\usepackage{accents}
\usepackage{subfigure}
\usepackage{enumitem}
\usepackage[dvipsnames]{xcolor}
\usepackage{hyperref}
\usepackage{ulem}

\hypersetup{
    colorlinks=true,
    linkcolor=blue,
    filecolor=magenta,
   citecolor=blue
}
\def\be{\begin{equation}}
\def\ee{\end{equation}}
\def\bea{\begin{eqnarray}}
\def\eea{\end{eqnarray}}
\def\mL{\mathcal{L}}
\def\mH{\mathcal{H}}

\def\mA{\mathcal{A}}

\def\mP{\mathcal{P}}

\def\mM{\mathcal{M}}

\def\zi{\mathrm{i}}
\def\ze{\mathrm{e}}

\def\zd{\mathrm{d}}

\def\zA{\mathrm{A}}
\def\zB{\mathrm{B}}

\def\ce{\varepsilon}

\def\hbmu{\bar{\mu}}
\def\hbn{\bar{\nu}}
\def\hbr{\bar{\rho}}
\def\hbs{\bar{\sigma}}

\def\hg{\hat{g}}
\def\hG{\hat{\Gamma}}
\def\hR{\hat{R}}
\def\hT{\hat{T}}
\def\hTe{\hat{\mathbb{T}}}
\def\he{\hat{e}}

\begin{document}
%\hfill  USTC-ICTS/PCFT-25-XX

\title{Correlated parity violation in gravity and electromagnetism from five-dimensional spacetime}

\author{Haomin Rao}
\email{rhm137@mail.ustc.edu.cn}
\affiliation{School of Intelligent Engineering, Shaoguan University, Shaoguan 512005, People's Republic of China}

\author{Yunlong Zheng$^{*}$}
\email{zhyunl@yzu.edu.cn (Corresponding author)}
\affiliation{Center for Gravitation and Cosmology, College of Physical Science and Technology, Yangzhou University, Yangzhou 225009, People's Republic of China}

\author{Qi-Zhe Hou}
\affiliation{School of Intelligent Engineering, Shaoguan University, Shaoguan 512005, People's Republic of China}

\author{Jian-Wen Ou}
\affiliation{School of Intelligent Engineering, Shaoguan University, Shaoguan 512005, People's Republic of China}

\author{Changyong Zhu}
\affiliation{School of Intelligent Engineering, Shaoguan University, Shaoguan 512005, People's Republic of China}

\begin{abstract}
Parity violation in the gravitational and electromagnetic sectors has been extensively investigated, yet the two are conventionally treated as independent phenomena. This separation, however, may be a four-dimensional prejudice. In higher-dimensional spacetime, gravity and electromagnetism may share a common geometric origin---and so, perhaps, does their parity violation. In this paper, we pursue this idea by constructing parity-violating Kaluza-Klein models in both Riemannian and teleparallel geometries. In Riemannian geometry, the simplest five-dimensional parity-violating term reduces to several complicated four-dimensional terms, including the familiar gravitational Chern-Simons term, which suffers from  ghost instability. In teleparallel geometry, however, the result is strikingly simple. The simplest five-dimensional parity-violating term reduces to only two ghost-free terms---the familiar Nieh-Yan term and the standard electromagnetic Chern-Simons  term. Remarkably, the model predicts that the helicity-dependent dispersion shift for electromagnetic waves is exactly six times that for gravitational waves, $\Delta\omega^2_{\rm EM}=6\,\Delta\omega^2_{\rm GW}$, a background-independent relation. This relation offers a falsifiable test of unification through joint cosmic microwave background and gravitational wave birefringence observations.
\end{abstract}

\maketitle

\section{Introduction}\label{sec:intro}

Motivated by recent observational advances in gravitational waves (GWs)
\cite{ligo1,ligo2} and the cosmic microwave background (CMB)
\cite{CMB1,CMB2}, parity-violating (PV) gravity theories have attracted
considerable interest in recent years. From the observational perspective, the
most distinctive signatures of parity-violating gravity are amplitude and
velocity birefringence---the left- and right-handed circularly polarized GWs
experience different amplitudes or propagation speeds~\cite{Alexander:2004us,Satoh:2007gn,Nishizawa:2018srh}.
These signals can be probed by ground-based GW detectors such as LIGO-Virgo-KAGRA~\cite{LVK:2021gwtc3,Zhao:2022CPT,Zhu:2023rrx}, as well as by pulsar timing arrays (PTAs)~\cite{NANOGrav:2023ng15,Fu:2023aab,Kato:2015bye,Belgacem:2020nda,Xu:2026dPTA}.
So far, various PV gravity models and their observable effects have been extensively studied in different gravitational frameworks~\cite{Qiao:2021fwi,Li:2021mdp,Cai:2021uup,Gong:2021jgg,Li:2022vtn,Zhang:2022xmm,Li:2022grj,Qiao:2022mln,Cai:2022lec,Chen:2022wtz,Zhu:2023lhv,Feng:2023veu,Lin:2023npr,Zhang:2023scq,Qiao:2023hlr,Li:2024fxy,Zhang:2024vfw,Akama:2024bav,Jiang:2024woi,Xiong:2024vsd,Xu:2024iao,Xu:2024kwy,Su:2025nkl,Li:2025bgu,Kang:2025rxn,Fu:2024ipa,Feng:2026pvst,Soda:2025cgwd,Song:2026pvscg,Horii:2025gf2f,Kang:2026quintom,Gao:2025qtg}.
On the other hand, parity violation has also been extensively investigated in the electromagnetic sector~\cite{Wilczek:1987mv,Carroll:1990vb,Harari:1992ea,Lue:1998mq,Li:2006ss,Li:2008tma,Li:2009rt,Komatsu:2022CMBreview}.
The most prominent example is cosmic birefringence, in which the electromagnetic
Chern-Simons (CS) term rotates the CMB polarization.
The Planck mission and its 2018 data release provide the broader CMB data
context~\cite{Planck:2018ov}.
Recent analyses of CMB polarization data have found tentative indications of isotropic cosmic birefringence~\cite{Minami:2020prl,Diego-Palazuelos:2022prl,Eskilt:2022prd}.
At present, however, parity violation in the gravitational and electromagnetic
sectors is conventionally treated as two independent phenomena, with unrelated
model parameters.

This separation is natural in four-dimensional spacetime: gravity and electromagnetism are described by independent degrees of freedom, $g_{\mu\nu}$ and $A_\mu$, with no geometric structure connecting them at the Lagrangian level. However, this independence may be a four-dimensional prejudice. The idea that spacetime has more than four dimensions has been a central theme in theoretical physics---from the original Kaluza-Klein proposal~\cite{Kaluza:1921tu,Klein:1926tv} to modern string theory~\cite{Green:1984sg,Duff:1985HeteroticKK}---and higher-dimensional unification remains one of the most compelling frameworks for understanding the relationship between fundamental forces. In the Kaluza-Klein (KK) theory~\cite{Duff:1986hr,Bailin:1987jd,Overduin:1997sri}, gravity and electromagnetism are unified in five-dimensional spacetime---$g_{\mu\nu}$ and $A_\mu$ are identified as different components of the five-dimensional metric $\hat{g}_{\bar{\mu}\bar{\nu}}$. If the five-dimensional theory contains a common PV interaction, then parity violation in the four-dimensional gravitational and electromagnetic sectors may be controlled by the same five-dimensional parameter, and the forms and relative coefficients of the resulting four-dimensional operators may consequently be correlated. To our knowledge, this unified perspective on parity violation has not been explored in the existing literature.

In this paper, we investigate how parity violation in the gravitational and electromagnetic sectors can be correlated through five-dimensional spacetime. We construct the simplest five-dimensional parity-violating action in the KK framework and dimensionally reduce it to four dimensions, analyzing the resulting parity-violating terms in both sectors. Importantly, we carry out this program in two geometric frameworks: Riemannian geometry and teleparallel geometry. The Riemannian framework is the natural starting point---the original KK theory is built upon it---but its parity-violating gravity sector is CS gravity~\cite{Jackiw:2003pm,Alexander:2009tp}, which contains higher-derivative terms and suffers from ghost instability~\cite{Dyda:2012rj}. The teleparallel framework, which describes gravity through torsion rather than curvature~\cite{Maluf:2013gaa,Bahamonde:2021gfp}, avoids higher derivatives in its PV sector and thus offers a promising route to a ghost-free theory.

Our results reveal a striking difference between the two frameworks.
In the Riemannian framework, the simplest five-dimensional PV term reduces to a complicated set of four-dimensional interactions.
The pure gravity sector is the CS gravity term~\cite{Jackiw:2003pm,Alexander:2009tp}, which inherits the well-known ghost instability.
The electromagnetic sector is even more intricate, involving the electromagnetic CS term together with gravity--electromagnetism mixing, quartic electromagnetic couplings, and derivative interactions with the KK scalar field.
In the teleparallel framework, however, the analogous construction reduces to only two four-dimensional terms---the Nieh-Yan term~\cite{Nieh:1981ww,Chandia:1997hu,Li:2020xjt,Li:2021wij} and the electromagnetic CS term $F\tilde{F}$~\cite{Wilczek:1987mv,Carroll:1990vb}.
No mixing, no extra couplings, and no higher derivatives appear; both terms are ghost-free.
Remarkably, we find that the helicity-dependent dispersion shift for electromagnetic waves is exactly six times that for gravitational waves, $\Delta\omega^2_{\rm EM}=6\,\Delta\omega^2_{\rm GW}$.
Note that this relation is exact and background-independent---it follows directly from the geometric structure of the five-dimensional theory.
It can be tested through joint observations of CMB birefringence and GW velocity birefringence, offering a concrete, falsifiable multi-messenger test of the unification hypothesis.

This paper is organized as follows.
We adopt a parallel structure, treating the Riemannian formulation first and the teleparallel formulation second.
Sec.~\ref{sec:kkriemann} reviews the standard Kaluza--Klein theory in Riemannian geometry, while Sec.~\ref{sec:pvriemann} constructs the five-dimensional PV term and performs its dimensional reduction.
Secs.~\ref{sec:kktg} and \ref{sec:pvtg} then develop the corresponding teleparallel construction and derive the $6\times$ relation between the electromagnetic and gravitational dispersion shifts.
Sec.~\ref{sec:pheno} discusses the resulting observational signatures, including CMB birefringence and GW velocity birefringence, and Sec.~\ref{sec:conclusion} summarizes our conclusions.
The fully covariant four-scalar construction and its zero-mode matching are introduced directly in Sec.~\ref{sec:pvriemann}. Appendix~\ref{app:diff-completion} gives further details, including the obstruction to describing a nontrivial KK direction by a single scalar.

In this paper, we adopt the metric signature $(-,+,+,+)$ and units in which $8\pi G=1$.
Five-dimensional spacetime indices are denoted by $\bar{\mu},\bar{\nu},\ldots=0,1,2,3,5$, and four-dimensional spacetime indices by $\mu,\nu,\ldots=0,1,2,3$.
The index ``5'' labels the extra spatial dimension.
Local Lorentz indices are denoted by $A,B,C,\ldots=0,1,2,3,5$ for five dimensions and $a,b,c,\ldots=0,1,2,3$ for four dimensions.

\section{Kaluza-Klein theory in Riemannian geometry}\label{sec:kkriemann}

In this section, we review the standard Kaluza-Klein theory in Riemannian geometry---the framework in which gravity and electromagnetism are unified as different manifestations of five-dimensional spacetime geometry~\cite{Duff:1986hr,Bailin:1987jd}.
We present a self-contained exposition of the essential geometric structure, establishing the notation and framework that will be used throughout this paper. For a comprehensive historical review, we refer the reader to Ref.~\cite{Overduin:1997sri}.

\subsection{1+4 decomposition of the five-dimensional metric}

We consider a five-dimensional spacetime manifold $M_{5}\cong M_{4}\times S^{1}$, where $M_{4}$ is a four-dimensional Lorentzian manifold and $S^{1}$ is a circle.
To perform the KK reduction, we choose the $U(1)$ circle action associated with the $S^{1}$ factor and denote its generator by $\zeta^{\hbmu}$. Here $\zeta^{\hbmu}$ only specifies the reduction direction and does not appear in the original five-dimensional Einstein--Hilbert action. We then choose an adapted coordinate system $\{x^{\hbmu},~\hbmu=0,1,2,3,5\}$ such that $\zeta=\partial/\partial x^{5}$. The $x^{5}$ coordinate lines---the curves along which $x^{\mu}={\rm const}$---therefore wind around the compact $S^{1}$ direction.
We denote the coordinate basis vectors by $\partial/\partial x^{\hbmu}$ and the dual basis one-forms by $\zd x^{\hbmu}$.

Let $\hg_{\hbmu\hbn}$ be the five-dimensional metric.
We introduce a normalized vector field $n^{\hbmu}$ tangent to the $x^{5}$ coordinate lines. The generator $\zeta^{\hbmu}=(\partial/\partial x^{5})^{\hbmu}$ is parallel to $n^{\hbmu}$, so that
\be\label{sec2:nu}
\zeta^{\hbmu}\equiv\left(\frac{\partial}{\partial x^{5}}\right)^{\hbmu}=\phi\, n^{\hbmu},
\qquad
\phi\equiv\left(\hg_{\hbmu\hbn}\zeta^{\hbmu}\zeta^{\hbn}\right)^{1/2},
\ee
where $\phi$ is a scalar field that characterizes the magnitude of the metric component along the fifth dimension, and $n^{\hbmu}=\zeta^{\hbmu}/\phi$ satisfies $n^{\hbmu}n_{\hbmu}=1$.
In analogy with the standard $1+3$ decomposition, although $(\partial/\partial x^{5})^{\hbmu}$ is parallel to $n^{\hbmu}$, the dual one-form $(\zd x^{5})_{\hbmu}$ is generally not parallel to $n_{\hbmu}$.
We thus write
\be\label{sec2:nd}
(\zd x^{5})_{\hbmu}=\phi^{-1}\,n_{\hbmu}-\sqrt{2}A_{\hbmu}~~\text{with}~~ n_{\hbmu}A^{\hbmu}=0,
\ee
where $A_{\hbmu}$ is the projection of $(\zd x^{5})_{\hbmu}$ onto the subspace orthogonal to $n^{\hbmu}$.
As we shall see, $A_{\hbmu}$ will play the role of the electromagnetic potential in four dimensions. The factor $\sqrt{2}$ is introduced to obtain the conventional coefficient for the electromagnetic kinetic term after dimensional reduction.
From Eqs.~(\ref{sec2:nu}) and (\ref{sec2:nd}), the coordinate components of $n^{\hbmu}$ and $n_{\hbmu}$ are
\be
n^{\hbmu}=(0^{\mu}, \phi^{-1}),\qquad n_{\hbmu}=(\sqrt{2}\,\phi A_{\mu}, \phi).
\ee
Let $g_{\hbmu\hbn}$ denote the metric projected onto the subspace orthogonal to $n^{\hbmu}$, satisfying $g_{\mu 5}=0$ and $g_{55}=0$.
The five-dimensional metric can then be decomposed as
\be\label{sec2:metric5}
\hg_{\hbmu\hbn}=g_{\hbmu\hbn}+n_{\hbmu}n_{\hbn}=
\left(
\begin{array}{cc}
g_{\mu\nu}+2\phi^{2}A_{\mu}A_{\nu} & \sqrt{2}\,\phi^{2}A_{\mu}\\
\sqrt{2}\,\phi^{2}A_{\mu} & \phi^{2}
\end{array}
\right).
\ee
The inverse metric is given by
\be\label{sec2:invmetric5}
\hg^{\hbmu\hbn}=
\left(
\begin{array}{cc}
g^{\mu\nu} & -\sqrt{2}A^{\mu}\\
-\sqrt{2}A^{\mu} & \phi^{-2}+2A_{\mu}A^{\mu}
\end{array}
\right),
\ee
where $g^{\mu\nu}$ is the inverse of $g_{\mu\nu}$ and $A^{\mu}=g^{\mu\nu}A_{\nu}$.
Eq.~(\ref{sec2:metric5}) is an exact geometric decomposition---no approximation has been introduced.
We have thus identified three four-dimensional fields from a single five-dimensional metric: a tensor field $g_{\mu\nu}$ (the four-dimensional metric), a vector field $A_{\mu}$ (the electromagnetic potential), and a scalar field $\phi$ (the dilaton, or radion).

To complete the reduction to four dimensions, the KK theory imposes a crucial physical assumption known as the cylinder condition~\cite{Kaluza:1921tu,Overduin:1997sri}:
the radius of the compact fifth dimension is so small that, at low energies, all physical fields are invariant under the circle action.
Quantization along the compact circle yields a Kaluza-Klein tower of massive modes whose masses are inversely proportional to the compactification radius; below the compactification scale, only the zero modes survive~\cite{Klein:1926tv,Duff:1986hr,Bailin:1987jd}.
In particular, $\mathcal{L}_{\zeta}\hg_{\hbmu\hbn}=0$; in the adapted coordinates introduced above, this is equivalent to $\partial_{5}\hg_{\hbmu\hbn}=0$, which leads to
$\partial_{5}\phi=0$, $\partial_{5}A_{\mu}=0$, and
$\partial_{5}g_{\mu\nu}=0$, so that $\phi$, $A_{\mu}$, and $g_{\mu\nu}$ can
be regarded as genuine four-dimensional fields.
This embodies the first key insight of the KK framework: higher-dimensional geometry provides a unified description of gravity and other fundamental interactions---the four-dimensional fields emerge as different components of a single, higher-dimensional metric.

\subsection{Gauge symmetry and the origin of the \texorpdfstring{$U(1)$}{U(1)} invariance}

The five-dimensional Einstein--Hilbert parent action of the original KK model does not explicitly contain $\zeta^{\hbmu}$ and is therefore invariant under the full five-dimensional diffeomorphism group. Once a circle action has been selected and the theory is restricted to the corresponding zero-mode sector, the diffeomorphisms that preserve this compactification structure descend to four-dimensional diffeomorphisms and the local $U(1)$ gauge symmetry. The former can be understood as the transformations that preserve the compact fifth-dimensional structure. We now focus on how the electromagnetic $U(1)$ gauge transformation arises from the fiber-preserving five-dimensional diffeomorphisms.

Let us consider a new coordinate system $\{\tilde{x}^{\hbmu}\}$ with $\tilde{x}^{\mu}=x^{\mu}$ and $\tilde{x}^{5}=\tilde{x}^{5}(x^{\mu},x^{5})$.
Since the $\tilde{x}^{5}$ coordinate lines are parallel to the original $x^{5}$ coordinate lines, we have the analogous decompositions
\be\label{sec2:newn}
\left(\frac{\partial}{\partial \tilde{x}^{5}}\right)^{\hbmu}=\tilde{\phi}\, n^{\hbmu},\qquad
(\zd \tilde{x}^{5})_{\hbmu}=\tilde{\phi}^{-1}\,n_{\hbmu}-\sqrt{2}\tilde{A}_{\hbmu}
\qquad\text{with}\qquad n_{\hbmu}\tilde{A}^{\hbmu}=0.
\ee
Combining Eqs.~(\ref{sec2:nu}), (\ref{sec2:nd}), and (\ref{sec2:newn}) with the tensor transformation law yields the relations between the old and new fields:
\be\label{sec2:relation0}
\tilde{\phi}=\phi\left(\frac{\partial \tilde{x}^{5}}{\partial x^{5}}\right)^{-1},\qquad
\tilde{A}_{\mu}=\left(\frac{\partial \tilde{x}^{5}}{\partial x^{5}}\right)A_{\mu}
-\frac{1}{\sqrt{2}}\frac{\partial \tilde{x}^{5}}{\partial x^{\mu}},\qquad
\tilde{g}_{\mu\nu}=g_{\mu\nu}.
\ee

Now consider the specific subclass of transformations
\be
\tilde{x}^{5}=x^{5}+\sqrt{2}\,\theta(x^{\mu}),
\ee
which corresponds to a redefinition of the $x^{5}=0$ hypersurface---a translation along the compact circle that varies from point to point in the four-dimensional spacetime.
Under this transformation, Eq.~(\ref{sec2:relation0}) reduces to
\be\label{sec2:U1}
\tilde{\phi}=\phi,\qquad
\tilde{A}_{\mu}=A_{\mu}-\partial_{\mu}\theta,\qquad
\tilde{g}_{\mu\nu}=g_{\mu\nu}.
\ee

Eq.~(\ref{sec2:U1}) is precisely the $U(1)$ gauge transformation of electromagnetism.
Note that the four-dimensional metric $g_{\mu\nu}$ is invariant under this transformation, confirming that it is indeed gauge-invariant in the electromagnetic sense; the scalar field $\phi$ is also invariant, establishing it as a gauge singlet.
This reveals the second key insight of the KK framework: the internal gauge symmetry of electromagnetism is interpreted as a geometric symmetry of the extra dimension---a manifestation of five-dimensional diffeomorphism invariance.

\subsection{The original Kaluza-Klein action and its dimensional reduction}

We now turn to the dynamics.
The original KK model is based on five-dimensional Riemannian geometry.
The Levi-Civita connection is
\be\label{sec2:connection}
\hG^{\hbr}_{~\hbmu\hbn}=\frac{1}{2}\,\hg^{\hbr\hbs}
\bigl(\partial_{\hbmu}\hg_{\hbn\hbs}+\partial_{\hbn}\hg_{\hbmu\hbs}-\partial_{\hbs}\hg_{\hbmu\hbn}\bigr),
\ee
and the Riemann curvature tensor is
\be\label{sec2:riemann}
\hR_{\hbmu\hbn\hbr}{}^{\hbs}=
\partial_{\hbn}\hG^{\hbs}_{~\hbmu\hbr}-\partial_{\hbmu}\hG^{\hbs}_{~\hbn\hbr}
+\hG^{\hbs}_{~\hbn\bar{\lambda}}\hG^{\bar{\lambda}}_{~\hbmu\hbr}
-\hG^{\hbs}_{~\hbmu\bar{\lambda}}\hG^{\bar{\lambda}}_{~\hbn\hbr}.
\ee
The five-dimensional Ricci tensor and curvature scalar
are $\hR_{\hbmu\hbn}=\hR_{\hbmu\hbr\hbn}{}^{\hbr}$ and $\hR=\hg^{\hbmu\hbn}\hR_{\hbmu\hbn}$, respectively.

The action of the original KK model is the five-dimensional Einstein-Hilbert action,
\be\label{sec2:action0}
S_{\rm KK}=\int \zd^{5}x\,\sqrt{-\hg}\,\frac{\hR}{2},
\ee
where $\hg=\det(\hg_{\hbmu\hbn})$.
Starting from the metric decomposition in Eq.~(\ref{sec2:metric5}), a straightforward but lengthy calculation yields
\be\label{sec2:reduction}
\sqrt{-\hg}=\phi\sqrt{-g},\qquad
\hR=R-\frac{1}{2}\phi^{2}F^{\mu\nu}F_{\mu\nu}+2\phi^{-1}\Box\phi,
\ee
where $F_{\mu\nu}=\partial_{\mu}A_{\nu}-\partial_{\nu}A_{\mu}$ is the electromagnetic field strength, $g=\det(g_{\mu\nu})$, $R$ is the four-dimensional Ricci scalar constructed from $g_{\mu\nu}$, and $\Box = -g^{\mu\nu}\nabla_{\mu}\nabla_{\nu}$ with $\nabla_{\mu}$ the covariant derivative compatible with $g_{\mu\nu}$.

Substituting Eq.~(\ref{sec2:reduction}) into the action in Eq.~(\ref{sec2:action0}) and integrating over the compact fifth dimension, we obtain the four-dimensional effective action:
\be\label{sec2:action1}
S_{\rm KK}=L_{5}\int \zd^{4}x\,\sqrt{-g}\left(\frac{\phi R}{2}-\frac{1}{4}\phi^{3}F^{\mu\nu}F_{\mu\nu}\right),
\ee
where $L_{5}=\int \zd x^{5}$ is the length of the fifth dimension, and we have discarded a total derivative term involving $\Box\phi$ that does not contribute to the equations of motion.

Varying the effective action in Eq.~(\ref{sec2:action1}) with respect to $\phi$, $A_{\mu}$, and $g_{\mu\nu}$ yields the four-dimensional field equations:
\bea\label{sec2:eom1}
& &\nonumber
\Box\phi=-\frac{1}{2}\phi^{3}F^{\mu\nu}F_{\mu\nu}
\\ & &\nonumber
\nabla^{\nu}F_{\nu\mu}=-3\phi^{-1}\nabla^{\nu}\phi\,F_{\nu\mu}
\\ & &
G_{\mu\nu}=\phi^{2}T^{\rm EM}_{\mu\nu}+\phi^{-1}\bigl(\Box\phi\,g_{\mu\nu}+\nabla_{\mu}\nabla_{\nu}\phi\bigr),
\eea
where $G_{\mu\nu}=R_{\mu\nu}-\frac{1}{2}g_{\mu\nu}R$ is the four-dimensional Einstein tensor and
$T^{\rm EM}_{\mu\nu}=F^{\rho}_{~\mu}F_{\rho\nu}-\frac{1}{4}F^{\rho\sigma}F_{\rho\sigma}g_{\mu\nu}$ is the standard electromagnetic energy-momentum tensor.

An equivalent and conceptually illuminating route to the field equations is to vary the original five-dimensional action in Eq.~(\ref{sec2:action0}) directly with respect to $\hg^{\hbmu\hbn}$, obtaining the five-dimensional vacuum Einstein equations:
\be\label{sec2:eom0}
\hat{G}_{\hbmu\hbn}=0,
\ee
where $\hat{G}_{\hbmu\hbn}=\hR_{\hbmu\hbn}-\frac{1}{2}\hg_{\hbmu\hbn}\hR$ is the five-dimensional Einstein tensor.
It can be seen that Eq.~(\ref{sec2:eom0}) encodes a profound result: the single five-dimensional vacuum equation $\hat{G}_{\hbmu\hbn}=0$, when decomposed into four-dimensional components, simultaneously yields the Einstein field equations (with the electromagnetic field as a source), the Maxwell equations (with a scalar-field coupling correction), and the equation of motion for the scalar field $\phi$.
In this sense, the KK framework achieves a genuine unification: general relativity (GR) and electromagnetism are not merely placed together by hand but emerge as inseparable aspects of pure five-dimensional geometry.

It is worth noting that the scalar field $\phi$ is not a passive spectator---its non-trivial dynamics modify the effective coupling strengths in four dimensions.
This can be seen directly from Eq.~(\ref{sec2:action1}): the gravitational sector couples as $\phi R$ rather than $R$ alone, and the electromagnetic sector couples as $\phi^{3}F^{2}$ rather than $F^{2}$.
At low energies, if $\phi$ stabilizes to a constant vacuum expectation value, one recovers standard Einstein-Maxwell theory with a fixed effective Newton constant and gauge coupling.
The stabilization of $\phi$ and the associated moduli-fixing problem are important topics in their own right~\cite{Overduin:1997sri}, though they lie beyond the scope of the present work.

\section{Parity violation in Riemannian Kaluza-Klein theory}\label{sec:pvriemann}

We now construct a PV model in the Riemannian KK framework. In four-dimensional
CS gravity, the PV term is built from the Pontryagin density
${}^{*}R^{\mu\nu\rho\sigma}R_{\mu\nu\rho\sigma}$~\cite{Jackiw:2003pm,Alexander:2009tp},
whose four Levi-Civita indices can be fully contracted with two curvature
tensors. The analogous construction is different in five dimensions. The
five-dimensional Levi-Civita tensor
$\hat{\ce}_{\hbmu\hbn\hbr\hbs\bar\lambda}
=\sqrt{-\hg}\epsilon_{\hbmu\hbn\hbr\hbs\bar\lambda}$ has five indices,
where $\epsilon_{\hbmu\hbn\hbr\hbs\bar\lambda}$ is the totally antisymmetric
symbol with $\epsilon_{01235}=1$, whereas curvature tensors carry an even
number of indices. The metric and curvature tensors alone therefore cannot
saturate all five indices, and an additional geometric object with one
five-dimensional spacetime index is required.

To saturate the remaining index without introducing a nondynamical background
vector, we introduce four scalar fields
$\{\psi^{a},\ a=0,1,2,3\}$ and define
\be\label{app:four-scalar-current}
J^{\hbmu}\equiv\frac{1}{4!}
\hat\ce^{\hbmu\hbn\hbr\hbs\bar\lambda}
\epsilon_{abcd}\partial_{\hbn}\psi^a\partial_{\hbr}\psi^b
\partial_{\hbs}\psi^c\partial_{\bar\lambda}\psi^d,
\qquad
u^{(\psi)}_{\hbmu}\equiv
\frac{J_{\hbmu}}{\sqrt{J_{\hbn}J^{\hbn}}},
\ee
on the spacelike-current branch $J_{\hbn}J^{\hbn}>0$. Here
$\epsilon_{abcd}$ is the four-dimensional totally antisymmetric symbol. Both
$J^{\hbmu}$ and $u^{(\psi)}_{\hbmu}$ are constructed from fields that are
varied in the action, and no preferred direction is specified in advance.
Throughout this paper we set $\mL_{\psi}=0$. The fields $\psi^{a}$ are
St\"uckelberg fields rather than four additional matter fields with independent
kinetic terms; they nevertheless enter the PV interactions through
$u^{(\psi)}_{\hbmu}$ and are varied together with the metric or tetrad.

Suppose the low-energy scalar configuration satisfies
\be\label{app:four-scalar-condition}
\partial_{5}\psi^{a}=0,
\qquad
\det\!\left(\partial_{\mu}\psi^{a}\right)\neq0.
\ee
The first condition is the cylinder condition, while the second requires the
four scalar gradients to be independent. Their gradients then span the
cotangent space of the four-dimensional base, so the dual current
$J^{\hbmu}$ points along the compact circle. With a consistent orientation,
Eq.~(\ref{app:four-scalar-condition}) gives
$u^{(\psi)}_{\hbmu}=n_{\hbmu}$.

Although we set $\mL_{\psi}=0$, the scalar fields $\psi^{a}$ still enter the
parent actions through $u^{(\psi)}_{\hbmu}$ and therefore have field equations.
Let
$\mathcal E^{\hbmu\hbn}=\frac{2}{\sqrt{-\hg}}\delta S/\delta\hg_{\hbmu\hbn}$ and
$\mathcal E_{a}=\frac{1}{\sqrt{-\hg}}\delta S/\delta\psi^{a}$. Five-dimensional
diffeomorphism invariance gives the Noether identity
\be\label{sec3:noether-identity}
\hat\nabla_{\hbmu}\mathcal E^{\hbmu}{}_{\hbn}
-\mathcal E_{a}\partial_{\hbn}\psi^{a}\equiv0.
\ee
When the gravitational equations hold, Eq.~(\ref{sec3:noether-identity}) reduces to
$\mathcal E_{a}\partial_{\hbn}\psi^{a}=0$. The independence of the four
gradients in Eq.~(\ref{app:four-scalar-condition}) then implies
$\mathcal E_{a}=0$, so the scalar equations are not independent. The same
argument applies to the teleparallel formulation. Hence, after varying the
parent actions with respect to all fields, we may impose
Eq.~(\ref{app:four-scalar-condition}) and use
$u^{(\psi)}_{\hbmu}=n_{\hbmu}$ without losing an independent equation. The
expressions containing $n^{\hbmu}$ below are understood in this zero-mode sense.

\subsection{Five-dimensional PV terms}

First, let us consider the parity-odd term linear in the curvature. There is
only one possible contraction,
\be\label{sec3:rgpv0}
\hat{\mP}_{1}=\hat{\ce}^{\hbmu\hbn\hbr\hbs\bar\lambda}
\hR_{\hbmu\hbn\hbr\hbs}u^{(\psi)}_{\bar\lambda}.
\ee
However, this term vanishes identically due to the Bianchi identity
$\hR_{[\hbmu\hbn\hbr]}{}^{\hbs}\equiv0$. Therefore, the simplest
nonvanishing parity-odd term must be quadratic in the curvature.

For the curvature-quadratic terms, the most symmetric contraction is
\be\label{sec3:rgpv1}
\hat{\mP}_{2}=\hat{\ce}^{\hbmu\hbn\hbr\hbs\bar\lambda}
\hR_{\hbmu\hbn\bar\alpha\bar\beta}
\hR_{\hbr\hbs}{}^{\bar\alpha\bar\beta}u^{(\psi)}_{\bar\lambda}.
\ee
It is worth noting that one may also construct other curvature-quadratic
parity-odd terms, such as
$\hat{\ce}^{\hbmu\hbn\hbr\hbs\bar\lambda}
\hR_{\hbmu\bar\alpha\hbn\bar\beta}
\hR_{\hbr\hbs}{}^{\bar\alpha\bar\beta}u^{(\psi)}_{\bar\lambda}$.
However, the Bianchi identity implies
$\hat{\ce}^{\hbmu\hbn\hbr\hbs\bar\lambda}
\hR_{\hbmu\bar\alpha\hbn\bar\beta}
\equiv\frac{1}{2}\hat{\ce}^{\hbmu\hbn\hbr\hbs\bar\lambda}
\hR_{\hbmu\hbn\bar\alpha\bar\beta}$. Therefore, this candidate is equal to
$\hat{\mP}_{2}/2$ and does not represent an independent operator. Consequently,
Eq.~(\ref{sec3:rgpv1}) is the only independent parity-odd term within the
curvature-quadratic class built from two Riemann tensors, one five-dimensional
Levi-Civita tensor, and $u^{(\psi)}_{\bar\lambda}$, without extra derivatives.
We thus obtain the fully covariant five-dimensional parent action
\be\label{sec3:rgcovaction}
S_{\rm R,cov}=\int\zd^{5}x\sqrt{-\hg}\left[
\frac{\hR}{2}
+c\hat{\ce}^{\hbmu\hbn\hbr\hbs\bar\lambda}
\hR_{\hbmu\hbn\bar\alpha\bar\beta}
\hR_{\hbr\hbs}{}^{\bar\alpha\bar\beta}u^{(\psi)}_{\bar\lambda}
\right],
\ee
where $c$ represents the strength of parity violation. After varying
Eq.~(\ref{sec3:rgcovaction}) and imposing the zero-mode conditions in
Eq.~(\ref{app:four-scalar-condition}), its parity-odd sector takes the
zero-mode form
\be\label{sec3:rgpvaction}
S_{\rm PV}=c\int\zd^{5}x\sqrt{-\hg}\,
\hat{\ce}^{\hbmu\hbn\hbr\hbs\bar\lambda}
\hR_{\hbmu\hbn\bar\alpha\bar\beta}
\hR_{\hbr\hbs}{}^{\bar\alpha\bar\beta}n_{\bar\lambda},
\ee
Equation~(\ref{sec3:rgpvaction}) is therefore used only as a convenient
representation of the covariant theory in its KK zero-mode sector. In
particular, the metric dependence inherited through
$u^{(\psi)}_{\hbmu}=n_{\hbmu}$ is retained when deriving the reduced field
equations.

\subsection{Dimensional reduction}

After varying the parent action and restricting to the zero-mode sector in
Eq.~(\ref{app:four-scalar-condition}), we may use
Eq.~(\ref{sec3:rgpvaction}) directly. Substituting the metric decomposition in
Eq.~(\ref{sec2:metric5}) and imposing the cylinder condition, we obtain, after a
lengthy calculation, the four-dimensional effective action
\be\label{sec3:rgpvaction1}
S_{\rm PV}=S_{\rm CSG}+S_{\rm PVEM}+S_{\rm mix}
+S_{\rm quarticEM}+S_{\rm otherEM},
\ee
where
\bea
& &\nonumber
S_{\rm CSG}=cL_{5}\int\zd^{4}x\sqrt{-g}\,
\phi\,{}^{*}R^{\mu\nu\rho\sigma}R_{\mu\nu\rho\sigma},
\\[2pt] & &\nonumber
S_{\rm PVEM}=2cL_{5}\int\zd^{4}x\sqrt{-g}\,
\phi X_{\phi}\,{}^{*}F^{\mu\nu}F_{\mu\nu},
\\[2pt] & &\nonumber
S_{\rm mix}=-4cL_{5}\int\zd^{4}x\sqrt{-g}\,
\phi^{3}\,{}^{*}R^{\mu\nu\rho\sigma}
\left(F_{\mu\nu}F_{\rho\sigma}+F_{\mu\rho}F_{\nu\sigma}\right),
\\[2pt] & &\nonumber
S_{\rm quarticEM}=4cL_{5}\int\zd^{4}x\sqrt{-g}\,
\phi^{5}\,{}^{*}F^{\mu\nu}F^{\rho\sigma}
\left(F_{\mu\nu}F_{\rho\sigma}+2F_{\mu\rho}F_{\nu\sigma}\right),
\\[2pt] & &
S_{\rm otherEM}=2cL_{5}\int\zd^{4}x\sqrt{-g}\,
\Big[\phi^{3}\nabla^{\lambda}{}^{*}F^{\mu\nu}\nabla_{\lambda}F_{\mu\nu}
+4\phi\nabla^{\lambda}{}^{*}F^{\mu\nu}
\left(\phi_{\mu}F_{\lambda\nu}+\phi_{\lambda}F_{\mu\nu}\right)
+8\phi_{\mu}\phi_{\lambda}F^{\lambda}_{~\nu}
{}^{*}F^{\mu\nu}\Big].
\label{sec3:reducedterms}
\eea
where $\phi_{\mu}=\nabla_{\mu}\phi$ and
$X_{\phi}=-g^{\mu\nu}\phi_{\mu}\phi_{\nu}$. Here
\begin{equation*}
{}^{*}F^{\mu\nu}=\frac{1}{2}\ce^{\mu\nu\rho\sigma}F_{\rho\sigma}
~~,~~
{}^{*}R^{\mu\nu\rho\sigma}
=\frac{1}{2}\ce^{\mu\nu\alpha\beta}R_{\alpha\beta}{}^{\rho\sigma}
\end{equation*}
are the Hodge duals of the electromagnetic field strength and the Riemann
tensor, respectively, and
$\ce_{\mu\nu\rho\sigma}=\sqrt{-g}\epsilon_{\mu\nu\rho\sigma}$ is the
four-dimensional Levi-Civita tensor.

It can be seen from Eq.~(\ref{sec3:rgpvaction1}) that the single
five-dimensional parity-odd term produces both gravitational and
electromagnetic PV terms. In addition, it produces mixed
gravity--electromagnetism interactions. Specifically, $S_{\rm CSG}$ is the
familiar CS gravity term, $S_{\rm PVEM}$ is the electromagnetic CS term,
$S_{\rm mix}$ couples the Riemann tensor to the
electromagnetic field, $S_{\rm quarticEM}$ is quartic in $F_{\mu\nu}$, and
$S_{\rm otherEM}$ contains derivatives of the electromagnetic field strength.
All these terms share the same coupling constant $c$, and their relative
coefficients are fixed by the five-dimensional parent action in
Eq.~(\ref{sec3:rgcovaction}). This means that, in the KK framework, parity
violation in the gravitational and electromagnetic sectors is not independent. This result
is a direct consequence of the KK unification: $g_{\mu\nu}$ and $A_{\mu}$ are
different components of the same five-dimensional metric, and thus both enter
the decomposition of the same five-dimensional curvature tensor.

\subsection{Tensor perturbations and ghost instability}

Among the terms in Eq.~(\ref{sec3:rgpvaction1}), $S_{\rm CSG}$ is precisely
the PV action of CS gravity, with the KK scalar $\phi$
playing the role of the CS coupling field. To see its cosmological effect and
stability, we consider an electromagnetic vacuum with $F_{\mu\nu}=0$. For
simplicity, we take the spatially flat Friedmann--Robertson--Walker (FRW)
background
$\zd s^{2}=a^{2}(\eta)(-\zd\eta^{2}+\delta_{ij}\zd x^{i}\zd x^{j})$,
where $a(\eta)$ is the scale factor, $\mH=a'/a$, and a prime denotes the
derivative with respect to the conformal time $\eta$. The Pontryagin density
vanishes on this homogeneous and isotropic background, so $S_{\rm CSG}$ does
not modify the background evolution. Its PV effect appears at
the perturbation level. We only consider the tensor perturbation
$h^{T}_{ij}$, so the perturbed metric is
\be\label{sec3:metricpert}
\zd s^{2}=a^{2}\left[-\zd\eta^{2}
+\left(\delta_{ij}+h^{T}_{ij}\right)\zd x^{i}\zd x^{j}\right].
\ee
We expand $h^{T}_{ij}$ in Fourier space and in terms of the left- and
right-handed circular polarization bases,
\be\label{sec3:Texpand}
h^{T}_{ij}(\eta,\vec{x})=
\sum_{\zA=L,R}\int\frac{\zd^{3}k}{(2\pi)^{3/2}}
h_{\zA}(\eta,\vec{k})\hat{e}_{ij}^{\zA}(\vec{k})
\ze^{\zi\vec{k}\cdot\vec{x}},
\ee
where the polarization bases satisfy
$\hat{e}_{ij}^{\zA}(\vec{k})\hat{e}_{ij}^{\zB}(\vec{k})=2\delta^{AB}$ and
$k_{l}\epsilon_{lik}\hat{e}_{jk}^{\zA}(\vec{k})
=\zi\mathfrak{p}_{\zA}k\hat{e}_{ij}^{\zA}(\vec{k})$, with
$\mathfrak{p}_{L}=-1$ and $\mathfrak{p}_{R}=+1$.

The quadratic action for the tensor perturbations is
\be\label{sec3:CStensoraction}
S^{(2)}_{TT}=L_{5}\sum_{\zA=L,R}\int\zd\eta\zd^{3}k\,
\frac{z_{\zA}^{2}}{4}
\left({h^{*}_{\zA}}'h_{\zA}'-k^{2}h^{*}_{\zA}h_{\zA}\right)
~~\text{with}~~
z_{\zA}^{2}=a^{2}\phi
\left[1+4\mathfrak{p}_{\zA}c(\ln\phi)'k/a^{2}\right].
\ee
Accordingly, the propagation equation for GWs is
\be\label{sec3:CStensoreom1}
h_{\zA}''+2\frac{z_{\zA}'}{z_{\zA}}h_{\zA}'+k^{2}h_{\zA}=0.
\ee
It can be seen that the friction term $2z_{\zA}'/z_{\zA}$ depends on the
helicity. Hence, the left- and right-handed GWs have different amplitude
evolutions, while the $k^{2}$ term and hence their phase velocities remain the
same. This is the amplitude birefringence phenomenon of CS gravity.

More importantly, if the action is extrapolated to sufficiently large $k$,
$z_{\zA}^{2}$ becomes negative for one of the two helicities. The kinetic
term of the corresponding mode then has a wrong sign, and this mode becomes a
ghost~\cite{Dyda:2012rj}. This is the well-known ghost instability of CS
gravity, which essentially originates from the higher derivatives
hidden in the curvature-quadratic CS interaction. Strictly speaking, CS
gravity may be treated as a low-energy effective theory with a cutoff
below the scale at which
$z_{\zA}^{2}$ changes sign~\cite{Alexander:2009tp,Zhao:2020PVwaveform}.
Nevertheless, the Riemannian model cannot be extrapolated to arbitrarily high
energies as a ghost-free theory.

The remaining terms in Eq.~(\ref{sec3:rgpvaction1}) contain the
electromagnetic field and do not contribute to the tensor perturbations on the
electromagnetic-vacuum background. Their cosmological effects deserve further
investigation. In principle, they can be studied by treating $A_{\mu}$ as a
first-order vector perturbation on the FRW background and deriving its
quadratic action. However, the ghost instability in the gravitational sector
already limits the theoretical viability of this construction. This motivates
us to seek a simpler and ghost-free realization of parity violation in the
teleparallel KK framework.

\section{Kaluza-Klein theory in teleparallel geometry}\label{sec:kktg}

Motivated by the ghost instability found in the Riemannian KK model,
we now turn to teleparallel geometry.
Related teleparallel approaches to KK unification have been discussed in Refs.~\cite{Andrade:2000KK,Barbosa:2002KK},
while several five-dimensional teleparallel models, their dimensional reductions, and related extensions have been considered in Refs.~\cite{Bamba:2013KK,Chang:2013WeylKK,Geng:2014FiveD,Geng:2014KK,Geng:2017KK}.
Here we develop a self-contained formulation tailored to the $1+4$ decomposition adopted in this work,
which will provide the geometric basis for the PV extension in the next section.
In teleparallel geometry, both
curvature and nonmetricity vanish, while gravity is described by torsion. In
four dimensions, the action constructed from the torsion scalar $\mathbb{T}$
is identical to the Einstein-Hilbert action up to a boundary term. The two
actions therefore give the same classical field equations. This formulation
is known as the teleparallel equivalent of general relativity (TEGR)~\cite{Maluf:2013gaa,Krssak:2018ywd}.
Beyond TEGR, various modified teleparallel theories, including
$f(\mathbb{T})$ gravity and new general relativity, have attracted
considerable attention~\cite{Ferraro:2006jd,Bengochea:2008gz,Linder:2010py,
Cai:2015emx,Hayashi:1979qx,Bahamonde:2017wwk,Bahamonde:2021gfp}.

The teleparallel version of the KK theory realizes the same unification in a
five-dimensional geometry with vanishing curvature and nonmetricity. After
dimensional reduction, the five-dimensional torsion gives rise to the
four-dimensional torsion, the electromagnetic field, and the KK scalar. Thus,
gravity and electromagnetism are again unified through the extra dimension,
with torsion replacing curvature as the fundamental geometric quantity.

In teleparallel geometry, one can choose an orthonormal tetrad, unique up to a
global Lorentz transformation, for which the spin connection vanishes. This
choice is referred to as the Weitzenb\"ock gauge~\cite{Krssak:2015oua,Krssak:2018ywd}. Since we adopt this gauge in
both five and four dimensions, we first need to determine the relation between
the corresponding tetrads. The KK decomposition singles out the normalized
vector $n^{\hbmu}$ along the compact direction. We therefore align the fifth
tetrad with this direction and take
$\he^{5}_{~\hbmu}=n_{\hbmu}=(\sqrt{2}\,\phi A_{\mu},\phi)$.
The remaining four tetrads are orthogonal to $n^{\hbmu}$ and take the form
$\he^{a}_{~\hbmu}=(e^{a}_{~\mu},0)$, where $\he^{A}_{~\hbmu}$ and
$e^{a}_{~\mu}$ denote the five- and four-dimensional tetrads, respectively.
The complete five-dimensional tetrad is therefore
\be\label{sec4:tetrad5}
\he^{A}_{~\hbmu}=
\left(
\begin{array}{cc}
e^{a}_{~\mu} & 0^{a}\\
\sqrt{2}\,\phi A_{\mu} & \phi
\end{array}
\right),
\ee
and its inverse is
\be\label{sec4:invtetrad5}
\he_{A}^{~\hbmu}=
\left(
\begin{array}{cc}
e_{a}^{~\mu} & -\sqrt{2}A_{a}\\
0^{\mu} & \phi^{-1}
\end{array}
\right),
\ee
where $e_{a}^{~\mu}$ is the inverse of $e^{a}_{~\mu}$ and
$A_{a}=e_{a}^{~\mu}A_{\mu}$. It can be verified that the metric
decomposition in Eq.~(\ref{sec2:metric5}) and the tetrad in
Eq.~(\ref{sec4:tetrad5}) satisfy
the orthonormality relation
\be\label{sec4:orthonormal}
\hg_{\hbmu\hbn}=\eta_{AB}\,\he^{A}_{~\hbmu}\,\he^{B}_{~\hbn}.
\ee
Thus, the tetrad decomposition contains the same four-dimensional fields
$e^{a}_{~\mu}$, $A_{\mu}$, and $\phi$ as the metric decomposition in
Sec.~\ref{sec:kkriemann}.

In the Weitzenb\"ock gauge, the five-dimensional torsion 2-form is
\be\label{sec4:torsion2form}
\hT^{A}_{~\hbmu\hbn}=\partial_{\hbmu}\he^{A}_{~\hbn}
-\partial_{\hbn}\he^{A}_{~\hbmu}.
\ee
Substituting Eq.~(\ref{sec4:tetrad5}) and imposing the cylinder condition, we
obtain
\bea\label{sec4:torsion5}
& &\nonumber
\hT^{a}_{~\mu\nu}=T^{a}_{~\mu\nu}
\\ & &\nonumber
\hT^{a}_{~\mu 5}=0~~,~~\hT^{a}_{~55}=0
\\ & &\nonumber
\hT^{5}_{~\mu\nu}=\sqrt{2}(\phi F_{\mu\nu}+2\nabla_{[\mu}\phi\,A_{\nu]})
\\ & &
\hT^{5}_{~\mu 5}=\nabla_{\mu}\phi~~,~~\hT^{5}_{~55}=0,
\eea
where $T^{a}_{~\mu\nu}=\partial_{\mu}e^{a}_{~\nu}
-\partial_{\nu}e^{a}_{~\mu}$ is the four-dimensional torsion 2-form and
$F_{\mu\nu}=\partial_{\mu}A_{\nu}-\partial_{\nu}A_{\mu}$ is the
electromagnetic field strength. It can be seen from
Eq.~(\ref{sec4:torsion5}) that the five-dimensional torsion decomposes into
the four-dimensional torsion $T^{a}_{~\mu\nu}$, the electromagnetic field
strength $F_{\mu\nu}$ together with its mixing with $A_{\mu}$, and the
derivative $\phi_{\mu}$ of the KK scalar. In particular, the electromagnetic
field appears as a component of the five-dimensional torsion in teleparallel
KK theory, just as it appears through the mixed components of the
five-dimensional metric in Riemannian KK theory.

Following the form of the four-dimensional torsion scalar, we introduce the
five-dimensional torsion scalar
\be\label{sec4:TS5}
\hat{\mathbb{T}}=-\frac{1}{4}\hT_{\hbmu\hbn\hbr}\hT^{\hbmu\hbn\hbr}
-\frac{1}{2}\hT_{\hbmu\hbn\hbr}\hT^{\hbr\hbn\hbmu}
+\hT^{\hbmu}\hT_{\hbmu}.
\ee
Here $\hT^{\hbr}_{~\hbmu\hbn}=\he_{A}^{~\hbr}
\hT^{A}_{~\hbmu\hbn}$ is the five-dimensional torsion tensor and
$\hT_{\hbmu}=\hT^{\hbr}_{~\hbmu\hbr}$ is the torsion vector. A direct
calculation gives the relation between the five- and four-dimensional torsion
scalars,
\be\label{sec4:TSrelation}
\hat{\mathbb{T}}=\mathbb{T}-\frac{1}{2}\phi^{2}F^{\mu\nu}F_{\mu\nu}
+2\phi^{-1}\phi_{\mu}T^{\mu},
\ee
where $\mathbb{T}$ and $T^{\mu}$ are the four-dimensional torsion scalar and
torsion vector, respectively.

The five-dimensional action of the teleparallel KK theory is
\be\label{sec4:tgaction0}
S_{\text{TKK}}=\int\zd^{5}x\,\sqrt{-\hg}\,
\frac{\hat{\mathbb{T}}}{2}.
\ee
Substituting Eq.~(\ref{sec4:TSrelation}) and integrating over the compact fifth
dimension, we obtain the four-dimensional effective action
\be\label{sec4:tgaction1}
S_{\text{TKK}}=L_{5}\int\zd^{4}x\,\sqrt{-g}
\left(\frac{\phi\mathbb{T}}{2}
-\frac{1}{4}\phi^{3}F^{\mu\nu}F_{\mu\nu}
+\phi_{\mu}T^{\mu}\right).
\ee
Using the identity $R=\mathbb{T}-2\nabla_{\mu}T^{\mu}$, where $R$ and
$\nabla_{\mu}$ are the Ricci scalar and the covariant derivative in
Riemannian geometry~\cite{Maluf:2013gaa,Krssak:2018ywd}, one can show that $S_{\text{TKK}}$ differs from the
Riemannian KK action $S_{\rm KK}$ only by a boundary term. Consequently, the
two actions yield the same classical field equations. Thus, the equivalence
between GR and TEGR also holds for their five-dimensional KK formulations.

This equivalence shows that the KK unification mechanism can be described by either curvature or torsion.
When the simplest
geometric scalar, $\hR$ or $\hat{\mathbb{T}}$, is adopted in five dimensions,
both formulations reduce to physically equivalent theories of gravity,
electromagnetism, and the KK scalar in four dimensions. In the next section,
we extend the teleparallel KK formulation by introducing a five-dimensional
PV term constructed from torsion.

\section{Parity violation in teleparallel Kaluza-Klein theory}\label{sec:pvtg}

Having established the teleparallel formulation of KK unification, we now
introduce parity violation through a five-dimensional torsion interaction. As
shown in Sec.~\ref{sec:pvriemann}, the curvature-based construction reduces to
several gravitational, electromagnetic, and mixed operators, while its CS
gravity sector suffers from a ghost instability. We will show that the
teleparallel construction leads instead to a considerably simpler
four-dimensional effective theory: a single torsion-quadratic PV term gives
the Nieh-Yan term~\cite{Nieh:1981ww,Chandia:1997hu,Li:2020xjt,Li:2021wij}, which constitutes the gravitational PV sector of the
Nieh-Yan modified teleparallel gravity model (hereafter the NYTG model), and
the standard electromagnetic CS term~\cite{Wilczek:1987mv,Carroll:1990vb}, with no mixed or quartic electromagnetic
operators.

\subsection{Five-dimensional PV term and dimensional reduction}

As in Sec.~\ref{sec:pvriemann}, the compact direction is supplied by the
St\"uckelberg combination $u^{(\psi)}_{\hbmu}$ in
Eq.~(\ref{app:four-scalar-current}), with $\mL_{\psi}=0$. The fundamental
five-dimensional action is therefore fully invariant under
$\mathrm{Diff}(M_{5})$. The unit vector $n^{\hbmu}$ will appear only after the
parent action has been varied and the full-rank zero-mode conditions in
Eq.~(\ref{app:four-scalar-condition}) have been imposed.

As in the Riemannian construction, a parity-odd scalar in five dimensions
requires the five-dimensional Levi-Civita tensor
$\hat{\ce}_{\hbmu\hbn\hbr\hbs\bar\lambda}
=\sqrt{-\hg}\epsilon_{\hbmu\hbn\hbr\hbs\bar\lambda}$. In teleparallel
geometry, the relevant building blocks are the torsion 2-form
$\hT^{A}_{~\hbmu\hbn}$ and the normalized vector
$u^{(\psi)}_{\hbmu}$, in addition to the metric. The local Lorentz indices of
two torsion 2-forms can be contracted with $\eta_{AB}$, whereas
$u^{(\psi)}_{\hbmu}$ saturates the remaining spacetime index of the
Levi-Civita tensor. The simplest and most symmetric torsion-quadratic parent
action constructed in this way is
\be\label{app:four-scalar-cov-action}
S_{\rm T,cov}=\int\zd^5x\sqrt{-\hg}\left[
\frac{\hTe}{2}
+\frac{c}{8}\eta_{AB}\hat\ce^{\hbmu\hbn\hbr\hbs\bar\lambda}
\hT^A_{~\hbmu\hbn}\hT^B_{~\hbr\hbs}u^{(\psi)}_{\bar\lambda}
\right].
\ee
After varying Eq.~(\ref{app:four-scalar-cov-action}) and imposing
Eq.~(\ref{app:four-scalar-condition}), its parity-odd sector becomes
\be\label{sec5:tgpvaction}
S_{\text{TPV}}=\frac{c}{8}\int\zd^{5}x\,\sqrt{-\hat{g}}\,\eta_{AB}\,\hat{\ce}^{\hbmu\hbn\hbr\hbs\bar\lambda}\,\hat{T}^{A}_{~\hbmu\hbn}\,\hat{T}^{B}_{~\hbr\hbs}\,n_{\bar\lambda},
\ee
where $c$ characterizes the strength of parity violation. Equation~(\ref{sec5:tgpvaction})
is the full-rank zero-mode form of the fully covariant parent action, not an
action containing a prescribed external direction. The numerical factor
$1/8$ is introduced only to facilitate comparison with previous studies of
the NYTG model~\cite{Li:2020xjt,Li:2021wij,Li:2022mti,Rao:2023doc}.

Within the zero-mode sector, substituting the torsion decomposition in
Eq.~(\ref{sec4:torsion5}) and imposing the cylinder condition gives the
four-dimensional effective action
\be\label{sec5:tgpvaction4d}
S_{\text{TPV}}=\frac{cL_{5}}{4}\int\zd^{4}x\,\sqrt{-g}\left(\phi\,\eta_{ab}\,{}^{*}T^{a\mu\nu}T^{b}_{~\mu\nu}+2\phi^{3}\,{}^{*}F^{\mu\nu}F_{\mu\nu}\right),
\ee
where ${}^{*}T^{a\mu\nu}=\frac{1}{2}\ce^{\mu\nu\rho\sigma}
T^{a}_{~\rho\sigma}$ is the Hodge dual of the torsion 2-form. The first term
in Eq.~(\ref{sec5:tgpvaction4d}) is precisely the Nieh-Yan term studied in the
NYTG model~\cite{Nieh:1981ww,Chandia:1997hu,Li:2020xjt,Li:2021wij}, whereas the second is the standard
electromagnetic CS term $\phi^{3}{}^{*}F^{\mu\nu}F_{\mu\nu}$~\cite{Wilczek:1987mv,Carroll:1990vb}.

This result is in sharp contrast to its Riemannian counterpart in
Eq.~(\ref{sec3:rgpvaction1}). There the dimensional reduction produces five types
of interactions, including mixed gravity--electromagnetism and quartic
electromagnetic terms. Eq.~(\ref{sec5:tgpvaction4d}), by contrast,
contains only two familiar PV operators. In particular, no mixed interaction
and no quartic electromagnetic term are generated. Nevertheless, the two
four-dimensional couplings are not independent: their relative coefficients
are fixed by the single five-dimensional parent action in
Eq.~(\ref{app:four-scalar-cov-action}).

\subsection{Tensor perturbations and gravitational-wave propagation}

We first consider tensor perturbations in an electromagnetic vacuum. Since the
metric does not fully determine the affine connection in teleparallel geometry,
we require both the metric and the connection to respect the cosmological
symmetries of spatial homogeneity and
isotropy~\cite{Hohmann:2019Symmetric,Hohmann:2020Cosmological,Coley:2022Symmetric}.
In the Weitzenb\"ock gauge, a convenient tetrad for the spatially flat FRW
background that satisfies these symmetry requirements can be chosen as
\be\label{sec5:backgroundtetrad}
\bar{e}^{0}_{~0}=a,\quad \bar{e}^{0}_{~i}=0,\quad \bar{e}^{a}_{~i}=a\delta^{a}_{i},
\ee
When only tensor perturbations are retained, the tetrad is expanded as \cite{Izumi:2012qj,Golovnev:2018wbh} :
\be\label{sec5:tensortetrad}
e^{0}_{~0}=a,\quad e^{0}_{~i}=0,\quad e^{a}_{~0}=0,\quad e^{a}_{~i}=a\,\delta^{a j}\bigl(\delta_{ij}+\frac{1}{2}h^{T}_{ij}\bigr),
\ee
where $h^{T}_{ij}$ denotes the tensor perturbation and satisfies the transverse
and traceless conditions. Expanding $h^{T}_{ij}$ in the helicity basis
according to Eq.~(\ref{sec3:Texpand}), the quadratic action for the tensor
perturbations of $S_{\text{TKK}}+S_{\text{TPV}}$ is
\be\label{sec5:Taction}
S^{(2)}_{TT}=L_{5}\sum_{\zA=L,R} \int \zd\eta\, \zd^{3}k\, \frac{a^{2}\phi}{4}\Big[ {h^{*}_{\zA}}'h_{\zA}'-\omega^{2}_{\zA}(k){h^{*}_{\zA}}h_{\zA}\Big]~~\text{with}~~\omega^{2}_{\zA}(k)=k^{2}\left[1+c\mathfrak{p}_{\zA}(\ln\phi)'/k\right],
\ee
and the corresponding propagation equation is
\be\label{sec5:Tpropagation}
h_{\zA}''+\left[2\mH+(\ln\phi)'\right]h_{\zA}'+\left[k^{2}+c\mathfrak{p}_{\zA}k(\ln\phi)'\right]h_{\zA}=0,
\ee
A general parameterization commonly used for modified GW propagation can be
written as
$h_{\zA}''+(2+\bar\nu+\nu_{\zA})\mH h_{\zA}'
+(1+\bar\mu+\mu_{\zA})k^{2}h_{\zA}=0$
~\cite{Nishizawa:2018srh,Zhu:2023rrx,Zhao:2020PVwaveform}. The present model belongs to the
subclass $\bar\mu=\nu_{\zA}=0$, with
\be\label{sec5:GWparameter_map}
\bar\nu=\frac{(\ln\phi)'}{\mH},\qquad \mu_{\zA}=\frac{c\mathfrak{p}_{\zA}(\ln\phi)'}{k}.
\ee
The wave-number dependence of $\mu_{\zA}$ corresponds to the
$\beta_{\mu}=-1$ PV dispersion induced by the Nieh-Yan term in the general
classification of Ref.~\cite{Zhu:2023rrx}. Comparing
Eq.~(\ref{sec5:Tpropagation}) with the standard propagation equation of the
NYTG model also identifies
$M_{\mathrm{PV}}(t)=c\,\zd\ln\phi/\zd t$ at the level of the propagation
equation~\cite{Wu:2021ndf}. The parameter $\bar\nu$, on the other hand, is a
helicity-independent friction correction induced by the evolution of the KK
scalar.

The helicity dependence of $\omega_{\zA}^{2}$ implies that the left- and
right-handed GWs have different phase evolutions and propagation velocities.
The model therefore predicts velocity birefringence, rather than the amplitude
birefringence found in the CS gravity sector of the Riemannian construction.
In the teleparallel PV KK model, the gravitational part of the reduced PV term
is precisely the Nieh-Yan term. Previous perturbative studies of the NYTG model
have shown that the Nieh-Yan term introduces no ghost modes in the scalar,
vector, or tensor sectors
~\cite{Li:2020xjt,Li:2021wij,Li:2022mti,Rao:2023doc}. The tensor-sector result
can also be seen directly from Eq.~(\ref{sec5:Taction}): the PV correction
introduces no higher-order time derivatives, and the kinetic coefficient
$a^{2}\phi/4$ is the same for both helicities and remains positive on the
physical branch $\phi>0$. This is in sharp contrast to the CS gravity sector of
the Riemannian construction.

\subsection{Electromagnetic perturbations and the sixfold dispersion relation}

The electromagnetic sector of Eq.~(\ref{sec5:tgpvaction4d}) is equally
transparent. Its PV interaction is the standard electromagnetic CS term,
whose birefringent effects have been extensively studied in
Refs.~\cite{Wilczek:1987mv,Carroll:1990vb,Harari:1992ea,Lue:1998mq,Li:2006ss,Li:2008tma,Li:2009rt,Komatsu:2022CMBreview}. To isolate this sector,
we keep only the flat FRW  background and regard the electromagnetic
field as a first-order perturbation. Taking $A_{0}=0$ and
$A_{i}=A^{V}_{i}$ , where the superscript $V$ denotes the transverse vector perturbation satisfying $\partial^{i}A^{V}_{i}=0$,
we expand its transverse vector mode in the circular
polarization basis as
\be\label{sec5:Vexpand}
A^{V}_{i}(\eta, \vec{x})=\sum_{\zA=L,R}\int \frac{\zd^{3}k}{(2\pi)^{\frac{3}{2}}}\, A_{\zA}(\eta, \vec{k})\, \hat{V}_{i}^{\zA}(\vec{k})\,\ze^{\zi\vec{k}\cdot\vec{x}},
\ee
where the polarization vectors satisfy
$\hat{V}_{i}^{\zA}(\vec{k})\hat{V}_{i}^{\zB}(\vec{k})=\delta^{AB}$ and
$\epsilon_{ijk}k_{j}\hat{V}_{k}^{\zA}(\vec{k})
=\zi\mathfrak{p}_{\zA}k\,\hat{V}_{i}^{\zA}(\vec{k})$ with
$\mathfrak{p}_{L}=-1$ and $\mathfrak{p}_{R}=+1$.
The quadratic action
for the electromagnetic vector perturbations is then
\be\label{sec5:Vaction}
S^{(2)}_{EM}=L_{5}\sum_{\zA=L,R} \int \zd\eta\, \zd^{3}k~ \phi^{3}\Big[ {A^{*}_{\zA}}'A_{\zA}'-\omega^{2}_{\zA}(k){A^{*}_{\zA}}A_{\zA}\Big]~~\text{with}~~\omega^{2}_{\zA}(k)=k^{2}\left[1+6c\mathfrak{p}_{\zA}(\ln\phi)'/k\right].
\ee

The standard electromagnetic CS term is likewise ghost-free. As can be seen
directly from Eq.~(\ref{sec5:Vaction}), it leaves the kinetic coefficient
$\phi^{3}$ unchanged and modifies only the helicity-dependent dispersion
relation. On the physical branch $\phi>0$, this coefficient is positive for
both helicities, so the electromagnetic sector contains no ghost mode.

Defining the helicity-dependent dispersion shift by
$\Delta\omega^{2}_{\zA}=\omega^{2}_{\zA}-k^{2}$, Eqs.~(\ref{sec5:Taction}) and
(\ref{sec5:Vaction}) give the exact relation
\be\label{sec5:dispersion_relation}
\Delta\omega^{2}_{\mathrm{EM}}=6\Delta\omega^{2}_{\mathrm{GW}}.
\ee
Thus, in the present field variables and at the level of the leading-order
Wentzel--Kramers--Brillouin (WKB) dispersion equations, the electromagnetic
dispersion shift is exactly six
times its GW counterpart. Importantly, Eq.~(\ref{sec5:dispersion_relation})
does not require any particular evolution of $a(\eta)$ or $\phi(\eta)$ and is
therefore independent of the specific cosmological background evolution. The
factor six is fixed entirely by the tensor contraction in the
five-dimensional parent action in Eq.~(\ref{app:four-scalar-cov-action}); it is an intrinsic relation
between the two propagation equations of the theory.

\subsection{Other PV torsion terms and their stability}

Although the PV term in Eq.~(\ref{app:four-scalar-cov-action}) is the most
symmetric torsion-quadratic contraction and yields the richest phenomenology among the operators considered
here, it is not the only parity-odd contraction that can be formed. Using the
five-dimensional Levi-Civita tensor, the torsion 2-form, the torsion vector
$\hat{T}_{\hbmu}$, and the St\"uckelberg vector
$u^{(\psi)}_{\bar\lambda}$, one may
construct three further terms,
\bea\label{sec5:PV_other_5d}
& &\nonumber
\mP_{2}=\hat{\ce}^{\hbmu\hbn\hbr\hbs\bar\lambda}\,\hat{T}_{\hbmu}\hat{T}_{\hbn\hbr\hbs}\,u^{(\psi)}_{\bar\lambda},
\\ & &\nonumber
\mP_{3}=\hat{\ce}^{\bar\lambda\hbmu\hbn\hbr\hbs}\,\hat{T}_{~\bar\alpha\hbmu\hbn}\hat{T}_{\hbr\hbs}{}^{\bar\alpha}\,u^{(\psi)}_{\bar\lambda},
\\ & &
\mP_{4}=\hat{\ce}^{\hbmu\hbn\hbr\hbs\bar\lambda}\,\hat{T}_{\hbmu\hbn\bar\alpha}\hat{T}_{\hbr\hbs}{}^{\bar\alpha}\,u^{(\psi)}_{\bar\lambda}.
\eea
Substituting the $1+4$ decomposition in Eq.~(\ref{sec4:torsion5}), these operators
reduce respectively to
\bea\label{sec5:PV_other_4d}
& &\nonumber
\mP_{2}=\ce^{\mu\nu\rho\sigma}T_{\mu}T_{\nu\rho\sigma}+\phi^{-1}\ce^{\mu\nu\rho\sigma}\nabla_{\mu}\phi\, T_{\nu\rho\sigma},
\\ & &\nonumber
\mP_{3}=\ce^{\mu\nu\rho\sigma}T^{\lambda}_{~\mu\nu}T_{\rho\sigma\lambda},
\\ & &
\mP_{4}=\ce^{\mu\nu\rho\sigma}T_{\mu\nu}{}^{\lambda}T_{\rho\sigma\lambda}.
\eea
It follows directly that $\mP_{2}$, $\mP_{3}$, and $\mP_{4}$ involve only
the four-dimensional torsion and its contractions; none of them contains the
electromagnetic field strength $F_{\mu\nu}$.
Their dimensional reductions therefore violate parity only in the gravitational sector, while leaving the electromagnetic sector parity symmetric.
This shows that, even when gravity and electromagnetism are unified in five dimensions,
a five-dimensional PV interaction need not induce parity violation simultaneously
in both sectors of the four-dimensional effective theory.

These alternative contractions, however, face a serious stability problem.
Cosmological perturbation analyses of general parity-odd interactions
quadratic in torsion have shown that, except for the Nieh-Yan term,
negative-kinetic modes can arise in the scalar and vector
sectors~\cite{Li:2022mti,Rao:2023doc}. More explicitly, when introduced as
standalone PV interactions, $\mP_{2}$, $\mP_{3}$, and $\mP_{4}$ each suffers
from a ghost instability in the scalar and/or vector perturbation sectors. They
therefore cannot individually define viable PV gravity models. This is the
reason why we restrict the main construction to the PV term in
Eq.~(\ref{app:four-scalar-cov-action}). Within the class of operators examined here, this
term is distinguished by three properties: the Nieh-Yan term in its
four-dimensional reduction avoids the above linear ghost problem, it violates
parity in both the gravitational and electromagnetic sectors, and the reduction
contains no interactions beyond the Nieh-Yan term and the standard
electromagnetic CS term.
The correlated observational consequences of the two helicity-dependent
dispersion relations will be investigated in the next section.

\section{Phenomenological implications}\label{sec:pheno}

In this section, we briefly discuss the leading observational signatures of the correlated electromagnetic and gravitational birefringence. A complete cosmological evolution and data analysis are beyond the scope of the present work.

We first consider the electromagnetic signal. For a small PV coupling $c$ or
a slowly varying KK scalar, the dispersion relation in
Eq.~(\ref{sec5:Vaction}) can be expanded as
$\omega_{\zA}(k)\approx k+3c\mathfrak{p}_{\zA}(\ln\phi)'$. The difference
between the phase velocities of the two circular polarizations is therefore
$(\omega_R-\omega_L)/k\approx6c(\ln\phi)'/k$. A linearly polarized
electromagnetic wave is a superposition of the two circular polarizations, and
their different phase evolutions rotate its polarization plane~\cite{Harari:1992ea,Lue:1998mq,Komatsu:2022CMBreview}. For CMB
photons propagating from the last-scattering surface (LSS) to the present time,
the accumulated rotation angle is
\be\label{sec6:rotation_angle}
\Delta\alpha(k,\eta_0)=\frac{1}{2}\int_{\eta_{\mathrm{LSS}}}^{\eta_0}\zd\eta\,\left[\omega_R(k,\eta)-\omega_L(k,\eta)\right]\approx3c\int_{\eta_{\mathrm{LSS}}}^{\eta_0}\zd\eta\,(\ln\phi)'=3c\ln\frac{\phi(\eta_0)}{\phi(\eta_{\mathrm{LSS}})}.
\ee
It can be seen that, at leading order, $\Delta\alpha$ is independent of
the wave number and is determined only by the evolution of the KK scalar
between the LSS and today. The rotation of the polarization plane mixes the
CMB Stokes parameters and, correspondingly, the $E$- and $B$-mode multipole
coefficients~\cite{Zaldarriaga:1996xe,Kamionkowski:1996ks}. Denoting the unrotated
coefficients by $a_{\ell m}^{E}$ and $a_{\ell m}^{B}$, their rotated
counterparts are
\be\label{sec6:eb_rotation}
\begin{pmatrix}
a_{\ell m}^{E'} \\[2pt]
a_{\ell m}^{B'}
\end{pmatrix}
=
\begin{pmatrix}
\cos(2\Delta\alpha) & -\sin(2\Delta\alpha) \\[2pt]
\sin(2\Delta\alpha) & ~~\cos(2\Delta\alpha)
\end{pmatrix}
\begin{pmatrix}
a_{\ell m}^{E} \\[2pt]
a_{\ell m}^{B}
\end{pmatrix}.
\ee
The resulting CMB angular power spectra are
\bea\label{sec6:cmb_rotation}
C_{\ell}^{EE'} &=& \cos^2(2\Delta\alpha)\,C_{\ell}^{EE} + \sin^2(2\Delta\alpha)\,C_{\ell}^{BB}, \nonumber\\
C_{\ell}^{BB'} &=& \sin^2(2\Delta\alpha)\,C_{\ell}^{EE} + \cos^2(2\Delta\alpha)\,C_{\ell}^{BB}, \nonumber\\
C_{\ell}^{EB'} &=& \frac{1}{2}\sin(4\Delta\alpha)\left(C_{\ell}^{EE} - C_{\ell}^{BB}\right), \nonumber\\
C_{\ell}^{TE'} &=& \cos(2\Delta\alpha)\,C_{\ell}^{TE}, \nonumber\\
C_{\ell}^{TB'} &=& \sin(2\Delta\alpha)\,C_{\ell}^{TE}.
\eea
In a statistically isotropic and parity-symmetric cosmology, the unrotated
cross-spectra $C_{\ell}^{EB}$ and $C_{\ell}^{TB}$ vanish. As shown explicitly
by Eq.~(\ref{sec6:cmb_rotation}), a homogeneous rotation generates both of
them and simultaneously redistributes power between the $E$ and $B$ modes.
Therefore, once Galactic foregrounds, polarization-angle calibration~\cite{Keating:2012ge}, and
other systematic effects are reliably controlled, a robust detection of
nonzero $C_{\ell}^{EB'}$ or $C_{\ell}^{TB'}$ would provide an observational
signature of parity violation
~\cite{Lue:1998mq,Minami:2020prl,Diego-Palazuelos:2022prl,Komatsu:2022CMBreview}.
In the present model, these spectra directly probe the combination
$c\ln[\phi(\eta_0)/\phi(\eta_{\mathrm{LSS}})]$. For example, the Planck PR4
analysis reported an isotropic rotation
$\Delta\alpha=0.30^{\circ}\pm0.11^{\circ}$ at the $68\%$ confidence level,
which corresponds in our convention to
$c\ln(\phi_0/\phi_{\mathrm{LSS}})=(1.75\pm0.64)\times10^{-3}$
~\cite{Diego-Palazuelos:2022prl}. This result remains sensitive to polarized
Galactic foregrounds and should not be regarded as conclusive evidence for
cosmological parity violation.

The GW sector provides a complementary probe. Expanding the dispersion
relation in Eq.~(\ref{sec5:Taction}) under the same approximation gives
$\omega_{\zA}(k)\approx k+\frac{1}{2}c\mathfrak{p}_{\zA}(\ln\phi)'$ and hence
$\Delta v_{\mathrm{GW}}\equiv(\omega_R-\omega_L)/k
\approx c(\ln\phi)'/k$. This is the velocity birefringence of GWs. For a
monochromatic mode emitted at $\eta_e$ and detected at $\eta_0$, the fixed-$k$
eikonal phase correction is
$\delta\Phi^{\rm WKB}_{\zA}=\int_{\eta_e}^{\eta_0}\zd\eta\,
[\omega_{\zA}(k)-k]=\frac{c}{2}\mathfrak{p}_{\zA}
\ln[\phi(\eta_0)/\phi(\eta_e)]$
~\cite{Zhao:2020PVwaveform,Mirshekari:2012MDR}. The phase difference between
the two helicities is therefore
\be\label{sec6:GW_phase_difference}
\Delta\Phi_{\rm GW}^{\rm WKB}\equiv\delta\Phi_R^{\rm WKB}-\delta\Phi_L^{\rm WKB}=c\ln\frac{\phi(\eta_0)}{\phi(\eta_e)}.
\ee
At leading order, this propagation phase difference is also independent of
the wave number. We emphasize that the fixed-$k$ eikonal phase is
not the stationary-phase Fourier-domain phase used in compact-binary
coalescence (CBC) analyses. Adopting the standard propagation-waveform
parametrization used in CBC analyses, the Fourier-domain correction for the
$\beta_{\mu}=-1$ case identified in Eq.~(\ref{sec5:GWparameter_map}) is written as
\be\label{sec6:CBC_phase}
\delta\Psi_{\zA}^{\rm CBC}(f)=\mathfrak{p}_{\zA}\mA_\mu\ln(\pi\mM f),\qquad \mA_\mu=\frac{1}{2}\int_{t_e}^{t_0}\zd t\,M_{\rm PV}(t)=\frac{c}{2}\ln\frac{\phi(t_0)}{\phi(t_e)},
\ee
where $\mM$ is the redshifted chirp mass
~\cite{Wu:2021ndf,Zhao:2020PVwaveform}. Existing analyses of compact-binary
signals have constrained the standard NYTG model and the general propagation
parametrization including $\beta_{\mu}=-1$, without finding robust evidence
for deviations from GR~\cite{Wu:2021ndf,Zhu:2023rrx}. However, the evolution
of $\phi$ in the present model also produces the common friction correction
$\bar\nu$ in Eq.~(\ref{sec5:GWparameter_map}). These constraints can therefore
serve only as qualitative references; a quantitative bound requires a
reanalysis with the complete waveform of the present model.

Finally, the electromagnetic and GW observables are fundamentally connected
by the exact and background-independent relation in
Eq.~(\ref{sec5:dispersion_relation}). Only when the two signals share the same
emission and detection times does its leading-order WKB integral reduce to
$\Delta\alpha=3\Delta\Phi_{\rm GW}^{\rm WKB}$, where the factor three, rather
than six, follows because the rotation angle is one half of the phase
difference between the two circular photon polarizations. This special
relation cannot be applied directly to CMB photons from the LSS at
$z\sim1100$ and compact-binary GWs from the low-redshift Universe with
$z\lesssim1$, since their propagation endpoints differ. It could instead be
tested with matched propagation histories, such as a GW event with a suitably
polarized electromagnetic counterpart. A joint analysis of CMB $B$ modes
sourced by primordial GWs and cosmic birefringence could provide a complementary
consistency test of the model, although it would not directly test the
same-endpoint relation.

\section{Conclusions}\label{sec:conclusion}

In this paper, we investigated how parity violation in the gravitational and
electromagnetic sectors can arise from a common geometric origin in
five-dimensional KK theory. We constructed the simplest five-dimensional PV
interaction in each of the Riemannian and teleparallel geometries and
dimensionally reduced both to four dimensions. In the Riemannian case, the curvature-quadratic interaction
produces the CS gravity term, the electromagnetic CS term,
gravity--electromagnetism mixing, quartic electromagnetic interactions, and
derivative couplings of the electromagnetic field. Their relative coefficients
are fixed by the same five-dimensional coupling, which means that the two PV
sectors are not independent. However, the CS gravity sector exhibits amplitude
birefringence and develops a ghost mode at sufficiently large wave numbers, so
this construction can at most be regarded as a low-energy effective theory below the corresponding cutoff and is not applicable in the large-wavenumber regime. In the teleparallel case, the
torsion-quadratic interaction gives a much simpler result: its dimensional
reduction contains only the Nieh-Yan term and the electromagnetic CS term,
without gravity--electromagnetism mixing or quartic electromagnetic
interactions. Studies of the NYTG model have shown that the Nieh-Yan term
introduces no ghost modes in the scalar, vector, or tensor perturbations, while
the electromagnetic CS term leaves the electromagnetic kinetic coefficient
positive on the physical branch $\phi>0$. Consequently, the resulting
teleparallel PV KK model is ghost-free. We also examined three other independent
torsion-quadratic PV contractions. Their reductions violate parity only in the
gravitational sector and, when introduced separately, suffer from ghost
instabilities in the scalar and/or vector perturbations. Thus, the
five-dimensional PV term adopted in our teleparallel PV KK model is the simplest
and most stable choice within the class of torsion-quadratic PV constructions
considered here. At the same time, its gravitational and electromagnetic
sectors exhibit helicity-dependent dispersion. At the level of the linear
propagation equations, their dispersion
shifts satisfy the exact and background-independent relation
$\Delta\omega^2_{\rm EM}=6\Delta\omega^2_{\rm GW}$, where the factor six is
fixed by the tensor contraction of the five-dimensional PV term. A more
detailed observational analysis based on specific CMB and GW data sets is left
for future work.

We emphasize that the fundamental five-dimensional actions in
Eqs.~(\ref{sec3:rgcovaction}) and
(\ref{app:four-scalar-cov-action}) are invariant under the full
$\mathrm{Diff}(M_{5})$ group. The compact direction is constructed from four
St\"uckelberg fields with $\mL_{\psi}=0$, and all fields are varied before the
full-rank KK zero-mode conditions are imposed. The Noether identity
in Eq.~(\ref{sec3:noether-identity}) ensures that no independent St\"uckelberg
equation is lost on the full-rank branch. The vector $n^{\hbmu}$ appearing in
the reduced calculation is consequently the zero-mode expression of
$u^{(\psi)\hbmu}$ rather than a preferred direction inserted into the parent
action. Appendix~\ref{app:diff-completion} gives further details of this
construction and explains why four scalar fields, rather than one, are needed
when $F_{\mu\nu}\neq0$.

\begin{acknowledgements}
The work of Haomin Rao was supported by the National Natural Science Foundation of China (NSFC) under Grant No. 12605117 and by the Scientific Research Startup Foundation of Shaoguan University under Grant No. 9900064901. The work of Yunlong Zheng was supported by the NSFC under Grant No. 11847239. The work of Qi-Zhe Hou was supported by the NSFC under Grant No. 12205194 and by the Science and Technology Planning Project of Shaoguan City under Grant No. 210809004530908. The work of Jian-Wen Ou was supported by the Guangdong Provincial Education Science Planning Project (Higher Education Special) under Grant No. 2024GXJK300. The work of Changyong Zhu was supported by the Talent Program of Shaoguan University under Grant No. 440-9900064502.
\end{acknowledgements}

\appendix
\section{Remarks on the tensor quadratic actions}
\label{app:matter-supported-frw}

In Secs.~\ref{sec:pvriemann} and \ref{sec:pvtg}, we directly presented the
quadratic actions for tensor perturbations without displaying their detailed
derivations. This appendix serves two purposes. First, we introduce an explicit
matter sector to provide a self-consistent FRW background and verify that the
tensor quadratic actions used in the main text are not restricted to a vacuum
 background. Second, we explain a convenient strategy for
organizing the calculation in the presence of parity-violating interactions.

To make the first point explicit, we introduce a canonical five-dimensional
scalar field $\varphi$ as a simple matter source. It is distinct from the KK
scalar $\phi$ and satisfies the cylinder condition
$\partial_{5}\varphi=0$. We derive the matter-corrected field equations and the
corresponding FRW background equations, and then use them to simplify the
parity-even tensor action. The Chern--Simons (CS) and Nieh--Yan (NY)
contributions can instead be evaluated separately and subsequently added to
the parity-even result. The purpose is to establish the background consistency
of the main-text results and clarify this computational simplification, rather
than to list every intermediate step of the perturbative expansion.

We begin with the parity-even Riemannian KK action supplemented by $\varphi$,
\be\label{app:matter-action-5d}
S_{\rm KK+\varphi}=\int \zd^{5}x\sqrt{-\hg}\left[
\frac{\hR}{2}-\frac{1}{2}\hg^{\hbmu\hbn}
\partial_{\hbmu}\varphi\partial_{\hbn}\varphi-V(\varphi)\right].
\ee
Using the metric decomposition in Eq.~(\ref{sec2:metric5}) and the cylinder
condition, this action reduces to
\be\label{app:matter-action-4d}
S_{\rm KK+\varphi}=L_{5}\int \zd^{4}x\sqrt{-g}\,\phi\left[
\frac{R}{2}-\frac{1}{4}\phi^{2}F^{\mu\nu}F_{\mu\nu}
-\frac{1}{2}g^{\mu\nu}\partial_{\mu}\varphi\partial_{\nu}\varphi
-V(\varphi)\right].
\ee
Varying Eq.~(\ref{app:matter-action-4d}) with respect to $\phi$, $A_{\mu}$,
$g_{\mu\nu}$, and $\varphi$, respectively, gives
\bea\label{app:matter-field-eqs}
& &\nonumber
\Box\phi=-\frac{1}{2}\phi^{3}F^{\mu\nu}F_{\mu\nu}
+\frac{2}{3}\phi V(\varphi),
\\ & &\nonumber
\nabla^{\nu}F_{\nu\mu}=-3\phi^{-1}\nabla^{\nu}\phi\,F_{\nu\mu},
\\ & &\nonumber
G_{\mu\nu}=\phi^{2}T^{\rm EM}_{\mu\nu}
+\phi^{-1}\bigl(\Box\phi\,g_{\mu\nu}
+\nabla_{\mu}\nabla_{\nu}\phi\bigr)+T^{\varphi}_{\mu\nu},
\\ & &
\Box\varphi-\phi^{-1}\nabla^{\mu}\phi\,\nabla_{\mu}\varphi
+V_{\varphi}=0.
\eea
Here
$T^{\rm EM}_{\mu\nu}=F^{\rho}_{~\mu}F_{\rho\nu}
-\frac{1}{4}F^{\rho\sigma}F_{\rho\sigma}g_{\mu\nu}$
is the standard electromagnetic energy-momentum tensor,
$T^{\varphi}_{\mu\nu}=\nabla_{\mu}\varphi\nabla_{\nu}\varphi
-\left(\frac{1}{2}\nabla^{\rho}\varphi\nabla_{\rho}\varphi
+V\right)g_{\mu\nu}$
is the scalar-field energy-momentum tensor, and
$V_{\varphi}=\zd V/\zd\varphi$.

We now take both $\phi$ and $\varphi$ to be homogeneous on the spatially flat
FRW background
$\zd s^{2}=a^{2}(\eta)(-\zd\eta^{2}+\delta_{ij}\zd x^{i}\zd x^{j})$
and set the background electromagnetic field to zero, $F_{\mu\nu}=0$.
Equation~(\ref{app:matter-field-eqs}) then yields
\bea\label{app:matter-background-eqs}
& &\nonumber
3\mH^{2}=-3\mH\frac{\phi'}{\phi}
+\frac{1}{2}{\varphi'}^{2}+a^{2}V,
\\ & &\nonumber
2\mH'+\mH^{2}=-\left(\frac{\phi''}{\phi}
+\mH\frac{\phi'}{\phi}\right)-\frac{1}{2}{\varphi'}^{2}+a^{2}V,
\\ & &\nonumber
\phi''+2\mH\phi'-\frac{2}{3}a^{2}\phi V=0,
\\ & &
\varphi''+\left(2\mH+\frac{\phi'}{\phi}\right)\varphi'
+a^{2}V_{\varphi}=0,
\eea
These equations provide an explicit matter-supported FRW
realization of the KK background.

On this matter-supported background, we now turn to tensor perturbations.
We expand the metric according to
Eqs.~(\ref{sec3:metricpert}) and (\ref{sec3:Texpand}). Scalar perturbations of
$\varphi$ do not mix with the transverse-traceless tensor modes at quadratic
order. Keeping terms up to second order and using
Eq.~(\ref{app:matter-background-eqs}), the parity-even tensor action becomes
\be\label{app:matter-tensor-action}
S^{(2)}_{TT,\rm even}=L_{5}\sum_{\zA=L,R}\int\zd\eta\,\zd^{3}k\,
\frac{a^{2}\phi}{4}\left(
{h^{*}_{\zA}}'h_{\zA}'-k^{2}h^{*}_{\zA}h_{\zA}\right).
\ee
The matter scalar $\varphi$ changes the background evolution but introduces no
additional kinetic or gradient term for the tensor perturbations. Equation
(\ref{app:matter-tensor-action}) is therefore precisely the parity-even limit
of Eqs.~(\ref{sec3:CStensoraction}) and (\ref{sec5:Taction}). For the
teleparallel construction, the same result follows because the parity-even
teleparallel and Riemannian KK actions differ only by a boundary term.

We next explain how the parity-violating contributions can be incorporated
without repeating the complete second-order calculation.
The full tensor quadratic action can be organized as
\be\label{app:tensor-action-split}
S^{(2)}_{TT}=S^{(2)}_{TT,\rm even}+S^{(2)}_{TT,\rm PV}.
\ee
The two parts can be treated differently. The parity-even part generally
contains terms that are simplified using the background equations. Although
the CS and NY interactions modify the field equations in a general spacetime,
their contributions vanish on the homogeneous and isotropic FRW background.
Therefore, the background equations remain those given in
Eq.~(\ref{app:matter-background-eqs}) and can be used to simplify
$S^{(2)}_{TT,\rm even}$. This does not imply that the parity-violating
contributions vanish at the perturbative level: the CS and NY interactions
still contribute nontrivially to the tensor action at quadratic order.

These contributions, collected in $S^{(2)}_{TT,\rm PV}$, can be obtained
directly by expanding the corresponding CS or NY interaction in tensor
perturbations. This calculation
does not rely on the detailed form of the background evolution equations.
Accordingly, the tensor quadratic actions for the two parity-violating
interactions considered here can be carried over from the existing results
without being rederived from scratch; only the coupling coefficients and
background quantities need to be translated into the conventions of the
present models.
Moreover, the matter scalar $\varphi$ does not enter either parity-odd
interaction directly; it affects the tensor actions only indirectly through
the background solutions for $a$ and $\phi$. This separation provides a
streamlined route to the full tensor quadratic action: the parity-even sector
is simplified with the matter-corrected background equations, while the
parity-violating contribution is calculated independently and then added.

For clarity, after the matter-corrected background equations are imposed, the
Riemannian result takes the form
\be\label{app:matter-CS-action}
S^{(2)}_{TT,\rm R}=L_{5}\sum_{\zA=L,R}\int\zd\eta\,\zd^{3}k\,
\frac{z_{\zA}^{2}}{4}
\left({h^{*}_{\zA}}'h_{\zA}'-k^{2}h^{*}_{\zA}h_{\zA}\right),
\qquad
z_{\zA}^{2}=a^{2}\phi
\left[1+4\mathfrak{p}_{\zA}c(\ln\phi)'k/a^{2}\right],
\ee
which is the result in Eq.~(\ref{sec3:CStensoraction}). In the teleparallel
construction, adding the NY contribution instead gives
\be\label{app:matter-NY-action}
S^{(2)}_{TT,\rm T}=L_{5}\sum_{\zA=L,R}\int\zd\eta\,\zd^{3}k\,
\frac{a^{2}\phi}{4}\left[
{h^{*}_{\zA}}'h_{\zA}'-\omega_{\zA}^{2}(k)h^{*}_{\zA}h_{\zA}\right],
\qquad
\omega_{\zA}^{2}(k)=k^{2}
\left[1+c\mathfrak{p}_{\zA}(\ln\phi)'/k\right],
\ee
in agreement with Eq.~(\ref{sec5:Taction}). This appendix therefore establishes
two complementary points. The explicit matter sector supplies a
self-consistent cosmological background for the tensor quadratic actions used
in the main text, showing that they are not restricted to vacuum background.
At the same time, the decomposition in
Eq.~(\ref{app:tensor-action-split}) provides a streamlined calculation of the
full quadratic action: the background equations are used to simplify the
parity-even sector, whereas the CS and NY contributions can be evaluated
separately and subsequently added.

\section{Further remarks on the covariant completion}
\label{app:diff-completion}

The main-text actions in Eqs.~(\ref{sec3:rgcovaction}) and
(\ref{app:four-scalar-cov-action}) construct the compact direction from the
four St\"uckelberg fields and preserve the full $\mathrm{Diff}(M_{5})$
symmetry. Their full-rank zero-mode forms in
Eqs.~(\ref{sec3:rgpvaction}) and (\ref{sec5:tgpvaction}) explicitly contain
$n^{\hbmu}$, but this vector is the zero-mode value of
$u^{(\psi)\hbmu}$ and is not an external structure of the parent actions. In
this appendix we explain why a single scalar cannot provide the required KK
direction when the electromagnetic field is nontrivial and give further
details of the four-scalar zero-mode sector.

A direct analogue of dynamical CS gravity would introduce a single scalar
$\vartheta$. On the spacelike-gradient branch
$X_{\vartheta}=\hg^{\hbmu\hbn}\partial_{\hbmu}\vartheta
\partial_{\hbn}\vartheta>0$, it defines the unit vector
\be\label{app:unit-vartheta}
u^{(\vartheta)}_{\hbmu}
=\frac{\partial_{\hbmu}\vartheta}{\sqrt{X_{\vartheta}}}.
\ee
Replacing $n_{\hbmu}$ by $u^{(\vartheta)}_{\hbmu}$ and supplying a scalar
Lagrangian would make the action fully invariant under five-dimensional
diffeomorphisms. This replacement, however, cannot reproduce the KK direction in the
presence of a nontrivial electromagnetic field. The vector in
Eq.~(\ref{app:unit-vartheta}) is identically hypersurface orthogonal and obeys
\be\label{app:single-scalar-frobenius}
\bigl(u^{(\vartheta)}\wedge\zd u^{(\vartheta)}\bigr)_{\hbmu\hbn\hbr}=0,
\ee
where $\zd$ denotes the exterior derivative and $\wedge$ the wedge product.
Equation~(\ref{sec2:nd}), on the other hand, implies
\be\label{app:kk-normal-frobenius}
(n\wedge\zd n)_{\hbmu\hbn\hbr}
=\sqrt{2}\phi^{2}\bigl[(\zd x^{5})_{\hbmu}+\sqrt{2}A_{\hbmu}\bigr]
\wedge F_{\hbn\hbr}.
\ee
Thus $u^{(\vartheta)}_{\hbmu}=n_{\hbmu}$ can hold only when the
electromagnetic field strength vanishes, $F_{\mu\nu}=0$. The KK direction is
hypersurface orthogonal only in the electromagnetic vacuum~\cite{Li:2023vvo},
whereas the normalized gradient of a scalar is always hypersurface orthogonal.
This kinematic obstruction cannot be removed by changing the scalar
Lagrangian, so a single scalar is insufficient for the theory studied in the
main text.

The obstruction is avoided by the four independent scalar fields introduced
in Eq.~(\ref{app:four-scalar-current}). Unlike a normalized scalar gradient,
$u^{(\psi)}_{\hbmu}$ is not constrained to be hypersurface orthogonal. It can
therefore reproduce the twisted KK direction even when $F_{\mu\nu}\neq0$.
The Riemannian and teleparallel parent actions are given explicitly in
Eqs.~(\ref{sec3:rgcovaction}) and
(\ref{app:four-scalar-cov-action}), respectively, with $\mL_{\psi}=0$ in both
cases.

Under the conditions in Eq.~(\ref{app:four-scalar-condition}), the four
gradients span the cotangent space of the four-dimensional base and the dual
current $J^{\hbmu}$ points along the compact circle. With a consistent
orientation, $u^{(\psi)}_{\hbmu}=n_{\hbmu}$. The parity-odd sectors of the two
parent actions then take the forms in Eqs.~(\ref{sec3:rgpvaction}) and
(\ref{sec5:tgpvaction}). The theory developed in the remaining main-text
calculation is thus the low-energy KK zero-mode sector of the covariant
construction.

The conditions in Eq.~(\ref{app:four-scalar-condition}) are mild and natural.
Because the fifth dimension is compact, modes carrying momentum along it are
heavy, so the cylinder condition also applies to the four scalars at low
energies. Their independence is an open, generic condition rather than an
additional field equation. The four-scalar construction can therefore recover
the KK direction without either exciting a nonzero KK mode or
imposing $F_{\mu\nu}=0$.

Because $\mL_{\psi}=0$, the St\"uckelberg fields have no independent kinetic
term and add no PV operator to the tensor or electromagnetic quadratic actions.
Both helicity-dependent dispersion shifts therefore remain unchanged, as does
the main-text relation
\be
\Delta\omega^2_{\rm EM}=6\Delta\omega^2_{\rm GW}.
\ee
More specifically, consider linear cosmological perturbations within the
low-energy KK zero-mode sector. If the four scalars, including their
perturbations, continue to satisfy Eq.~(\ref{app:four-scalar-condition}), then
the perturbations $\delta\psi^a$ can change only the magnitude of $J^{\hbmu}$,
not its direction. Consequently, the identity
$u^{(\psi)}_{\hbmu}=n_{\hbmu}$ continues to hold at the perturbative level. The
scalar zero-mode perturbations do not enter the reduced four-dimensional
effective action and therefore introduce no additional propagating degrees of
freedom. Their equations are not independent, as follows from the Noether
identity in Eq.~(\ref{sec3:noether-identity}).

The four-scalar construction therefore provides a parent theory that is fully
invariant under five-dimensional diffeomorphisms, reproduces the main-text
effective theory, and preserves
$\Delta\omega^2_{\rm EM}=6\Delta\omega^2_{\rm GW}$ under the stated low-energy
conditions. It is not intended as a unique microscopic completion; rather, it
demonstrates explicitly that the full-rank zero-mode theory follows from a
covariant dynamical embedding.

\end{document}